\documentclass[12pt]{article}
\usepackage{graphicx,a4}   
\usepackage{amssymb}        
\def\math{\mathsurround 0pt}
\def\oversim#1#2{\lower.5pt\vbox{\baselineskip0pt \lineskip-.5pt
        \ialign{$\math#1\hfil##\hfil$\crcr#2\crcr{\scriptstyle\sim}\crcr}}}
\def\lap{\mathrel{\mathpalette\oversim {\scriptstyle <}}}
 
\begin{document}
\newcommand{\preprintno}[1]
{{\normalsize\begin{flushright}#1\end{flushright}}}

\title{\preprintno{{\bf SUSX-TH/01-005}\\ hep-ph/0101131}Cosmological Constraints on Large Extra Dimensions}
\author{Malcolm Fairbairn\thanks{E-mail address: mdsf@star.cpes.sussex.ac.uk}\\
{\em Centre for Theoretical Physics/Astronomy Centre}\\
{\em University of Sussex,} \\
        {\em Brighton BN1 9QH, U.K.}}
\date{12th January 2001}

\maketitle
\begin{abstract}
We calculate the production of massive Kaluza-Klein(KK) modes via nucleon-nucleon gravi-bremsstrahlung in the early universe.  Overproduction of these states would result in early matter domination and therefore a low age for the universe so it is possible to place constraints on the number and size of large extra dimensions.  The constraints are stronger than those from Sn1987a - for 2 large extra dimensions and $T_{QCD}=$170 MeV, we show the fundamental scale must be larger than 1,000 TeV.

\end{abstract}

\section{Introduction}
Over the past few years various authors have considered the possibility of large extra dimensions as a solution to the discrepancy between the $\sim$ TeV scale of standard model physics and the apparent mass scale associated with gravitational interactions $10^{19}$GeV$=G^{-\frac{1}{2}}$.  In these models, the standard model particles are confined to a 3+1 dimensional brane whilst gravity is free to inhabit all 3+1+n dimensions.  The expression for the observed value of G in four dimensions is given by $\cite{arkani}$
\begin{equation}
4\pi G=\left(\frac{1}{RM_{f}}\right)^{n}\frac{1}{M_{f}^{2}}
\label{bigg}
\end{equation}
and so it is possible to obtain arbitrarily small values of the higher dimensional fundamental scale $M_{f}$ provided the radii $R$ of the higher dimensions are large enough.  

Factorisable higher dimensions (for example higher dimensions compactified on $T^{n}$ or $S^{n}$) have associated with them an infinite tower of massive KK states which can be produced during high energy scattering of standard model particles on the brane, a phenomenon which may be investigated at the LHC $\cite{ahern}$.  The possibility of this KK mode generation has already led to strong constraints on the size and number of higher dimensions from Astro-Particle Physics.  As one of the mechanisms expected to be responsible for the generation of the KK modes is gravi-bremsstrahlung from nucleon-nucleon scattering, the core of a supernova would seem to be a good place to look for the effect.  However, the observed neutrino pulse from Supernova 1987a successfully accounted for most of the energy of the collapse.  It was therefore possible to place constraints on the energy lost via this mechanism to KK modes and consequently the radius and number of the extra dimensions $\cite{sn,hanhart}$.  One would expect the same mechanism to apply to nucleon-nucleon scattering in the early universe\footnote{Although unlike a supernova core it may not be the dominant mechanism $\cite{hall}$}. 

The KK modes corresponding to phenomenologically interesting values of the size of the extra dimensions have lifetimes larger than the age of the universe $\cite{lykken}$ and dissipate as matter, i.e. $\rho_{KK}\propto R^{-3}$.  This extra injection of matter leads to an earlier matter-radiation equality which in turn leads to a more rapid decline in the CMB temperature.  The increased cooling rate means that by the time the CMB has cooled to 2.73K the universe is still be much too young to hold the objects we observe in ours.  It is these constraints that are investigated in this paper.

\section{Nucleon-Nucleon Gravi-Bremsstrahlung}
In $\cite{hanhart}$ the angle averaged emmissivity of the plasma into KK modes $\epsilon_{KK}$ from a non-degenerate gas of nucleons is shown to be given by the expression
\begin{equation}
\frac{d\epsilon_{KK}}{dt}=\frac{16 G_{4}}{9}\sqrt{\frac{T^{7}}{m_{N}}}(RT)^{n}n_{N}^{2}\sigma_{0}^{NN}C_{n}
\label{hanhart}
\label{emiss}
\end{equation}
Where T is temperature, $m_{N}$ is the isospin averaged nucleon mass $938.9$ MeV, $C_{n}$ is given in the appendix, and the radius of the large extra dimensions R is given by rearranging equation ($\ref{bigg}$)
\begin{equation}
R=(M_{f}^{2+n}4\pi G)^{-\frac{1}{n}}.
\end{equation}
In ($\ref{hanhart}$) the nucleon-nucleon cross section $\sigma_{0}^{NN}$ is evaluated at the peak of the centre of mass energy distribution $\propto\tau^{n+4}e^{-\tau}$ where $\tau=E_{com}/T$ so the cross section is evaluated at $E_{com}=(n+4)T$.  
 The total S-wave nucleon-nucleon scattering cross section is given by $\cite{wilson}$ 
\begin{equation}
\sigma_{0}^{NN}=\frac{4\pi}{k^{2}+(\frac{k^{2}\rho}{2}-\frac{1}{a^{2}})^{2}}
\end{equation}
where $a$ is the scattering length and $\rho$ is the effective range.  We can obtain approximate values for these parameters by using those experimentally observed for the neutron-neutron $^{1}S_{0}$ channel, $a=-18.5\pm 0.4 fm$ and $\rho=2.80\pm 0.11 fm$ \cite{teramond}. Now we can express the cross section at the peak of the energy distribution in MeV$^{-2}$
\begin{equation}
\sigma_{0}^{nn}=\frac{4\pi}{(n+4)^{2}T^{2}+\left[(n+4)^{2}T^{2}/(2.25\times 10^{4}\rm{MeV})+67\rm{MeV}\right]^{2}}
\end{equation}
where T is measured in MeV.  At this stage it is worth noting that although the cross section decreases with temperature by a factor of about 2 between 1 MeV and 200 MeV we will still expect the emissivity to be very temperature dependant because of equation (\ref{emiss}).

\section{Cosmology}
In order to ensure that we are correct in using the normal 4 dimensional FRW equations for our cosmology, we need to make sure that the density of matter in the bulk remains lower than the fundamental Planck Density $\rho_{(4+n)}=M^{4+n}_{f}$ by the factor $M_{f}^{2}G \ll 1$ $\cite{linde}$.  In this paper where we start with no KK modes, the total density in the bulk is given by the density lost from the brane divided by the volume of large extra dimensions.  Assuming a toroidal compactification manifold these constraints reduce to
\begin{equation}
\rho_{KK}<M_{f}^{n+6}G(2\pi R)^{n}=\frac{(2\pi)^{n-1}}{2}M_{f}^{4}.
\end{equation}
This inequality is easily satisfied throughout our calculations.

Secondly, the rate at which radiation energy density on the brane will evaporate into KK modes in the bulk was obtained in $\cite{arkani}$ by dimensional analysis.  One can then obtain a temperature at which the cooling of radiation is dominated by this evaporation into the bulk as opposed to the normal cooling due to cosmological expansion.  This temperature, $T_{*}$, serves as a maximum temperature up to which the universe can be considered 'normal'.  Comparing cooling rates: 
\begin{equation}
T_{*} \lap \left(\frac{M^{n+2}_{f}}{M_{pl}}\right)^{\frac{1}{n+1}}\sim10^{\frac{(6n-9)}{(n+1)}}\rm{MeV}\times\left(\frac{M_{f}}{1\rm{TeV}}\right)^{\frac{n+2}{n+1}}
\end{equation}
We are only considering events which occur below 170 MeV and the case with one large extra dimension has been ruled out by gravitational experiments for all but very high values of $M_{f}$.  Consequently, only the case with 2 large extra dimensions has a value of $T_{*}$ below 170 MeV and then only for values of $M_{f}$ below approximately 8 TeV which have been ruled out by the supernova work $\cite{sn,hanhart}$. Having checked the above two validity arguments we can continue with the analysis.

The energy conservation equation for the KK modes in an expanding universe is
\begin{equation}
\frac{d\rho_{KK}}{dt}=\frac{d\epsilon_{KK}}{dt}-3H\rho_{KK}.
\end{equation}
To convert this into a differential equation with respect to temperature, we use the standard expression for the radiation dominated epoch,
\begin{equation}
t=\frac{0.301}{\sqrt{g_{*}G_{4}}T^{2}}; \qquad H=1.66\sqrt{g_{*}G}T^{2}
\label{time}
\end{equation}
where $g_{*}$ is the total number of relativistic degrees of freedom in the plasma $\cite{kolb}$.  These degrees of freedom are given in the appendix. The radiation energy density is written
\begin{equation}
\rho_{\gamma}=\frac{g_{*}\pi^{2}}{30}T^{4}
\end{equation}
Then we can write
\begin{equation}
\frac{d\rho_{KK}}{dT}=-\frac{0.602}{\sqrt{g_{*}G}T}\frac{d\epsilon_{KK}}{dt}+\frac{3\rho_{KK}}{T}
\end{equation}
 The temperature range over which we shall consider the production of KK modes will be from the end of the QCD phase transition (170 MeV) downwards.  The temperature sensitivity of the KK mode generation rate means the vast majority of any mode production occuring in this range will take place at temperatures just below $T_{QCD}$, or whatever upper temperature we run the equations from.  

Because the detailed calculations of $\cite{hanhart}$ have been made for the nucleon-nucleon system, we should only use them once the QCD phase transition is over.  That is not to say that there will be no KK mode production by gravi-bremsstrahlung before the phase transition in the quark-gluon plasma\footnote{See $\cite{arndt}$ for a first approach to this calculation}, merely that that has not been explicitly calculated at the time of publication and the cosmology of a universe with large extra dimensions becomes increasingly uncertain as one pushes back towards higher energies.  The constraints obtained in this paper will assume that there are no KK modes excited at $T_{QCD}$, only those generated below this temperature will be considered.  Inclusion of any KK modes generated above $T_{QCD}$ would only strengthen the constraints obtained below. 

The QCD phase transition is well below the nucleon mass threshold so the neutrons and protons are non-relativistic and the distribution Boltzmann.
\begin{equation}
n_{N}=g\left(\frac{m_{N}T}{2\pi}\right)^{\frac{3}{2}}e^{-\frac{(m_{N}-\mu)}{T}}\qquad (m>>T)
\end{equation}
Here g is the number of degrees of freedom, 4 for a nucleon, and $m_{N}$ is the mass of the nucleon.  Any chemical potential corresponding to baryon number conservation will already be small and will decrease with the expansion so will not be included in these calculations.   

Finally there is a multiplicity factor $g_{m}=3$ which takes into account the fact that both n-n and n-p scattering will give rise to KK modes.

Bringing these results together, the rate of energy density generated in the form of KK modes with respect to temperature $\ref{emiss}$ will be given by
\begin{equation}
\frac{d\rho_{KK}}{dT}=-4.31\times 10^{-3}\frac{C_{n}g^{2}g_{m}}{\sqrt{g_{*}G}}\frac{m_{N}^{\frac{5}{2}}}{M_{(4+n)}^{2+n}}\frac{T^{\frac{11}{2}+n}e^{-\frac{2(m_{N}-\mu_{N})}{T}}}{(n+4)^{2}T^{2}+\left(\frac{(N+4)^{2}T^{2}}{2.25\times 10^{4}}+67\rm{MeV}\right)^{2}}+\frac{3\rho_{KK}}{T}
\end{equation}
\begin{figure}[tb]
\centering
\includegraphics[width=10.5cm,height=8cm]{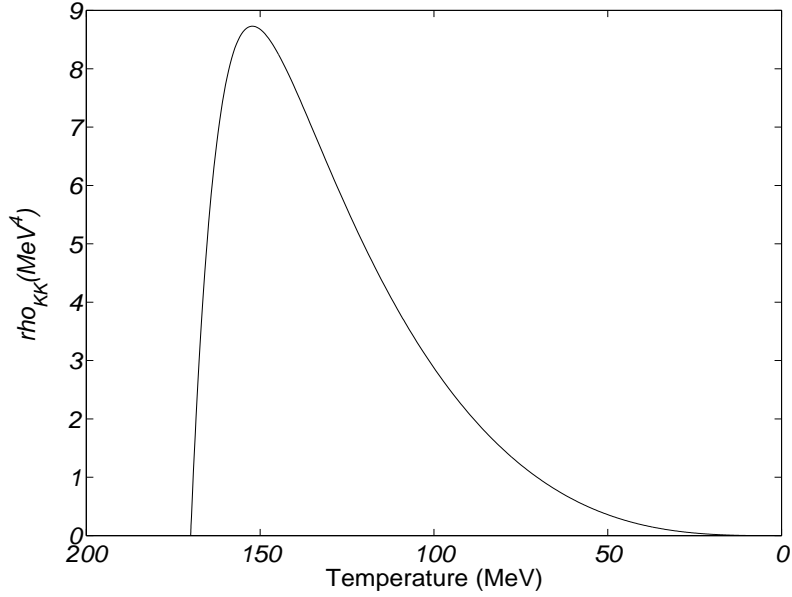}
\caption{Diagram showing a typical density profile of KK modes as the universe cools}
\end{figure}
The ratio of time vs. the epoch of radiation-matter equality $t_{eq}$ is $\cite{kolb}$
\begin{equation}
\frac{t}{t_{eq}}=\frac{(a/a_{eq}-2)(a/a_{eq}+1)^{\frac{1}{2}}+2}{2-\sqrt{2}}
\end{equation}
and the temperature of radiation $T_{\gamma}$ is inversely proportional to the scale factor $a$.  We know that the temperature of the cosmic microwave radiation today is equal to $2.73K=2.35\times 10^{-10}$ MeV.  If we  also assume that the minimum age of the universe is $12.8 Gyrs=4.1\times 10^{17}s=6.2\times 10^{39}\rm{MeV}^{-1}$ as determined by the mean observed age of globular clusters $\cite{krauss}$ then by using equation ($\ref{time}$) we can obtain a maximum temperature for $T_{eq}$, the temperature at radiation-matter equality.
\begin{equation} 
T_{eq}\le 2.7\times 10^{-6}MeV
\end{equation}
This is the limit we have to ensure is not broken by an early $t_{eq}$ due to overproduction of KK modes.
\section{Results and Discussion}
\begin{figure}[tb]
\centering
\includegraphics[width=10.5cm,height=8cm]{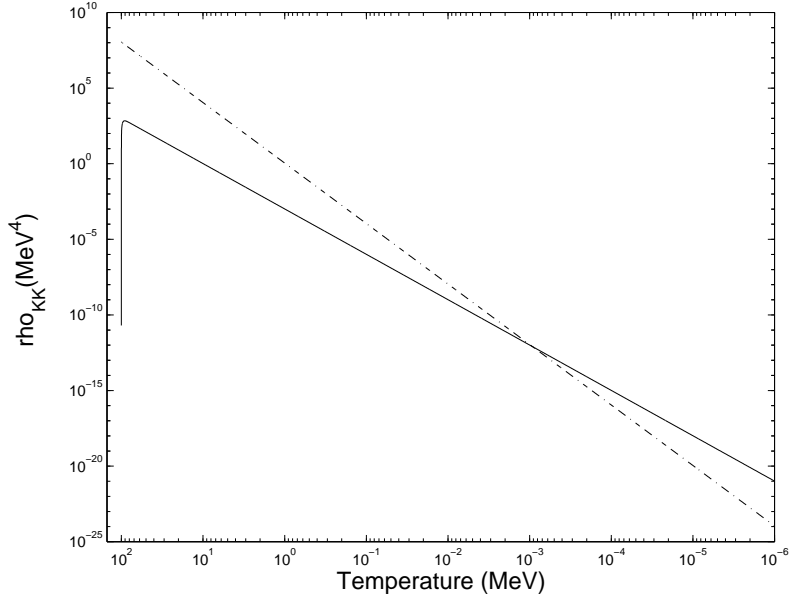}
\caption{Solid line is density of KK modes for n=2, $M_{f}$=20TeV, $T_{QCD}$=100MeV.  Dot-dash line is radiation density.  This example would correspond to $t_{today}\approx 10^{6}yrs$.}
\label{fig2}
\end{figure}

Limits on the minimum value of $M_{f}$ obtained for n=2, 3 and 4 for values of $T_{QCD}$=100, 135 and 170 MeV are shown in table $\ref{results}$.  These values of $T_{QCD}$ are quite low and therefore conservative as a higher $T_{QCD}$ rapidly increases the number of modes produced.  An example of a situation previously permitted by Sn1987a calculations was $n=2, M_{f}=20$ TeV.  As can be seen in figure \ref{fig2} this is no longer permitted even if one assumes that $T_{QCD}$ is as low as 100 MeV as it leads to an unacceptably high $T_{eq}$.  If one allows the KK mode production to begin at 170 MeV then $M_{f}$ must be greater than $10^{3}$ TeV for n=2.  This is no longer a solution of the hierarchy problem.

Even if one were to claim that the calculations in this paper were only valid for temperatures far below the QCD phase transition there would still be many other mechanisms by which KK modes could be produced via the bremsstrahlung or annihilation processes of other particles $\cite{hall}$.  

It is possible that these constraints could be avoided by some new cosmological evolution which occurs above 1 MeV to repect nucleosynthesis but at a temperature below about 100 MeV to avoid overproduction of KK modes.  Until that is achieved gravi-bremsstrahlung processes in the context of large extra dimensions appear to be at odds with cosmology.   

\begin{table*}
\begin{center}
\caption[]{\label{results} Minimum value of $M_{f}$ to prevent $t_{today}<12.8 Gyr$ for various $T_{QCD}$}
\begin{tabular}[tb]{|c|c|c|c|}	\hline  
Number	& &&\\ 
of extra 	&$T_{QCD}$=170MeV&$T_{QCD}$=135MeV&$T_{QCD}$=100MeV\\ 
dimensions	&&&\\\hline \hline
2		&1,000TeV&390TeV	&86TeV	\\ \hline
3		&59TeV	&26TeV	&7.4TeV	\\ \hline
4		&9.0TeV	&4.4TeV	&1.5TeV	\\ \hline
\end{tabular}
\end{center}
\end{table*}
\section*{Acknowledgments}
Thanks to David Bailin, Ed Copeland, Thomas Dent, Louise Griffiths, Steen Hannestad, Mark Hindmarsh, Sam Leach and Andrew Liddle for useful conversations.  The author is funded by PPARC.
\section*{Appendix}
The values of $C_{n}$ used in the text come from $\cite{hanhart}$
\begin{equation}
C_{n}=\pi^{\frac{n-1}{2}}h_{n}\frac{\Gamma(n+1)\Gamma(n+5)}{\Gamma(n+\frac{5}{2})}\frac{n^{2}+9n+23}{(2n+7)(2n+5)}
\end{equation}
where if we only consider the dominant emission to gravitons, the expression for $h_{n}$ is given by
\begin{equation}
h_{n}=\frac{3n^{2}+18n+19}{(n+5)(n+3)}\frac{1}{\Gamma(\frac{n+3}{2})}
\end{equation}
values for these parameters are given in table $\ref{gammas}$.
\begin{table*}
\begin{center}
\caption{\label{gammas} Different values of $h_{n}$ and $C_{n}$ for the dominant graviton emission into the bulk}
\begin{tabular}{|c|c|c|}	\hline  
n	&$h_{n}$			&$C_{n}$	\\ \hline
1	&$\frac{5}{3}$			&31.523		\\ \hline
2	&$\frac{268}{105\sqrt{\pi}}$ 	&143.63		\\ \hline
3	&$\frac{25}{24}$		&248.30$\pi$	\\ \hline
4	&$\frac{1112}{945\sqrt{\pi}}$	&1521.3$\pi$	\\ \hline
5	&$\frac{23}{60}$		&3253.4$\pi^{2}$\\ \hline
6	&$\frac{752}{2079\sqrt{\pi}}$   &23558$\pi^{2}$ \\ \hline
7	&$\frac{73}{720}$		&57853$\pi^{3}$ \\ \hline
\end{tabular}
\end{center}
\end{table*}
The temperatures used for each of the freeze out temperatures were obtained from $\cite{peacock}$ and are listed in table $\ref{freeze}$.
\begin{table*}
\begin{center}
\caption[]{\label{freeze} Degrees of freedom of the plasma at different temperatures}
\begin{tabular}{|c|c|c|}\hline
Particle	&Freezeout Temperature 	&$g_{*}$ after freezeout \\ \hline
$T_{QCD}$ (gluons)&170 MeV		&14.25  \\ \hline 
$\mu^{\pm}$ 	&109 MeV		&10.75	\\ \hline		
$\nu$		&$\approx$ 4.5 MeV	&6.86	\\ \hline
$e^{\pm}$ 	&4.3 MeV		&3.36   \\ \hline
\end{tabular}
\end{center}
\end{table*}

\end{document}